\documentclass[12pt]{article}
\pdfoutput=1

\newlength{\abstractwidth}
\usepackage{amsmath}
\usepackage{amsfonts}
\usepackage{amssymb}
\usepackage{latexsym}

\usepackage{epsf}
\usepackage{color}
\usepackage{graphicx}
\usepackage{tikz}
\usepackage{dsfont}
\usepackage{subfigure}

\usepackage{hyperref}
\definecolor{darkred}{rgb}{0.8,0.1,0.1}
\hypersetup{colorlinks=true, linkcolor=darkred, citecolor=blue, linktoc=page}

\usetikzlibrary{arrows,shapes,positioning}
\usetikzlibrary{decorations.markings}
\usepackage[rightcaption]{sidecap}
\tikzstyle arrowstyle=[scale=1]
\tikzstyle directed=[postaction={decorate,decoration={markings,
    mark=at position .65 with {\arrow[arrowstyle]{stealth}}}}]
\tikzstyle reverse directed=[postaction={decorate,decoration={markings,
    mark=at position .65 with {\arrowreversed[arrowstyle]{stealth};}}}]
\usetikzlibrary{positioning}

\renewcommand{\thefootnote}{\fnsymbol{footnote}}
\renewcommand{\thanks}[1]{\footnote{#1}}
\newcommand{\starttext}{
\setcounter{footnote}{0}
\renewcommand{\thefootnote}{\arabic{footnote}}}

\newcommand{\bea}{\begin{eqnarray}}
\newcommand{\eea}{\end{eqnarray}}
\newcommand{\be}{\begin{eqnarray}}
\newcommand{\ee}{\end{eqnarray}}

\def\cL{{\cal L}}
\def\cM{{\cal M}}
\def\cN{{\cal N}}

\def\RR{{\mathbb R}}

\def\half{{1\over 2}}
\def\p{\partial}

\def\a{\alpha}
\def\b{\beta}
\def\g{\gamma}

\def\eps{\epsilon}
\def\f{\varphi}

\def\d{\delta}
\def\s{\sigma}
\def\m{\mu}
\def\){\right)}
\def\({\left( }
\def\]{\right] }
\def\[{\left[ }

\def\no{\nonumber}

\makeatletter\def\l@subsubsection#1#2{}%
\makeatother

\begin{document}
\starttext
\setcounter{footnote}{0}

\vskip 0.3in

\begin{center}

{\Large \bf Janus solutions in six-dimensional gauged supergravity}

\vskip 0.4in

{\large   Michael Gutperle$^a$, Justin Kaidi$^a$  and Himanshu Raj$^b$} 

\vskip .2in

$^a$ {\it Mani L. Bhaumik Institute for Theoretical Physics}\\
{ \it Department of Physics and Astronomy }\\
{\it University of California, Los Angeles, CA 90095, USA}\\[0.5cm]

$^b${\it  SISSA and INFN} \\ 
 {\it Via Bonomea 265; I 34136 Trieste, Italy} \\
 
 \vskip .2in
 
\begin{abstract}

Motivated by an analysis of the sub-superalgebras of the five-dimensional superconformal algebra $F(4)$, we search for the holographic duals to co-dimension one superconformal defects in 5d CFTs which have $SO(4,2) \oplus U(1)$ bosonic symmetry. In particular, we look for domain wall solutions to six-dimensional $F(4)$ gauged supergravity coupled to a single vector multiplet. It is found that supersymmetric domain wall solutions do not exist unless there is a non-trivial profile for one of the vector multiplet scalars which is charged under the gauged $SU(2)$ R-symmetry. This non-trivial profile breaks the $SU(2)$ to $U(1)$, thus matching expectations from the superalgebra analysis. A consistent set of BPS equations is then obtained and solved numerically. While the numerical solutions are generically singular and thought to be dual to boundary CFTs, it is found that for certain fine-tuned choices of parameters regular Janus solutions may be obtained.

\end{abstract}
\end{center}

\baselineskip=16pt
\setcounter{equation}{0}
\setcounter{footnote}{0}

\newpage
\tableofcontents
\newpage
\section{Introduction}
\setcounter{equation}{0}
\label{sec1}

The study of five-dimensional supersymmetric field theories utilizing string theory was initiated in \cite{Seiberg:1996bd,Morrison:1996xf,Intriligator:1997pq}. Subsequently, many five-dimensional gauge theories have been realized using brane constructions in type IIA  \cite{Brandhuber:1999np,Bergman:2012kr} as well as type IIB string theory \cite{Aharony:1997ju,Aharony:1997bh,DeWolfe:1999hj}.  In the realm of supersymmetric field theories, superconformal theories (SCFTs)  take a special place. The case of five-dimensional  SCFTs is particularly interesting since there exists a unique superalgebra with $SO(2,5)$ conformal symmetry, $SU(2)$ R-symmetry, and sixteen supercharges (eight Poincare and eight special conformal supercharges). Together they form the superalgebra $F(4)$ \cite{Nahm:1977tg,Kac:1977em,Shnider:1988wh}. It is an interesting fact that in five dimensions the maximal superconformal theory has sixteen supercharges, unlike in dimensions three, four, and  six, where the maximal number of supercharges is thirty-two.

Many of the known five-dimensional field theories allow for limits where the rank of the gauge group(s) can be taken to be large. Consequently, a holographic dual of the superconformal phase of such theories should exist. Solutions containing an $AdS_6$ factor had previously been found in massive IIA supergravity \cite{Brandhuber:1999np,Bergman:2012kr,Passias:2012vp} as well as in type IIB supergravity \cite{Lozano:2012au,Lozano:2013oma,Kelekci:2014ima}.  However, all of these solutions suffer from singularities, where the supergravity approximation breaks down.

Recently,\footnote{For earlier work in this direction see \cite{Apruzzi:2014qva,Kim:2015hya,Kim:2016rhs}.} in a series of papers \cite{DHoker:2016ujz,DHoker:2016ysh,DHoker:2017mds}, new solutions of type IIB supergravity were constructed using a warped product of $AdS_6\times S^2$ over a two-dimensional Riemann surface $\Sigma$ with boundary. These solutions are completely regular away from isolated points on the boundary of $\Sigma$. The poles have a clear physical interpretation as the remnants of semi-infinite $(p,q)$ five-branes. These branes can be interpreted as the semi-infinite external five-branes which are used to construct five-dimensional field theories using $(p,q)$ five-brane webs  \cite{Aharony:1997ju,Aharony:1997bh,DeWolfe:1999hj}. The singularities do not affect the calculation of some holographic observables, such as the entanglement entropy of a spherical region and the free energy on a sphere \cite{Gutperle:2017tjo}. However, the fact that the solutions are a warped product of $AdS_6\times S^2$ over a two-dimensional surface makes it very hard to determine the full Kaluza-Klein spectrum of fluctuations or to calculate holographic correlation functions. 

A simpler setting for $AdS_6/CFT_5$ duality is given by six-dimensional $F(4)$ gauged supergravity. $F(4)$ gauged supergravity was first constructed in \cite{Romans:1985tw}. The theory can be coupled to six-dimensional vector multiplets and the general Lagrangian, supersymmetry transformations, and possible gaugings can be found in \cite{Andrianopoli:2001rs}. These theories have supersymmetric $AdS_6$ vacua, and determining the spectrum of linearized supergravity fluctuations dual to primary operators as well as correlation functions is straightforward \cite{Ferrara:1998gv,DAuria:2000afl,Karndumri:2016ruc}. For some additional work on the use of $F(4)$ gauged supergravity in holography, see e.g. \cite{Karndumri:2012vh,Alday:2014rxa,Alday:2014fsa,Hama:2014iea}.

Apart from local operators, a CFT in general also contains extended defect operators, such as Wilson lines, surface operators, domain walls, and interfaces. A $p$-dimensional defect in a $d$-dimensional CFT is called conformal when it is $SO(2,p)$ invariant, i.e. when it preserves the conformal transformations acting on its $p$-dimensional worldvolume. A special class of conformal defects are the so-called  superconformal defects, which preserve some fraction of the supersymmetry of the SCFT as well. This implies that the preserved symmetries of such a defect form  a superalgebra. Hence a classification of possible superconformal defects in a SCFT amounts to finding all the sub-superalgebras lying inside the original superalgebra of the SCFT with a bosonic $SO(2,p)$ factor. Such an analysis was undertaken for the maximal superalgebras with 32 supercharges and defects with 16 supercharges (so-called half-BPS defects) in \cite{DHoker:2008wvd}. 

For five-dimensional SCFTs the superalgebra is the particular real form of $F(4)$ which has $SO(5,2)\oplus SU(2)_R$ as its bosonic symmetry algebra. The sixteen supercharges transform in the ${\bf 8} \otimes {\bf 2}$  representation under the bosonic symmetry group. Happily, the sub-superalgebras of $F(4)$ have been classified in \cite{jeugt:1987,Frappat:2000} and are given along with their relevant real forms in the following table.

\begin{table}[htp]
\caption{Sub-superalgebras of $F(4)$}
\begin{center}
\begin{tabular}{|c|c|c|c|}
\hline
sub-superalgebra&  even part & even part real form&  susys\\
\hline
  $A_1\oplus B_3$ &  $A_1\oplus B_3$ & $ SO(5,2)\oplus SO(3)$ & 0\\
\hline
  $ A_2 \oplus A(0,1)$ &$A_2 \oplus A_1 \oplus U(1) $& $SO(2,1) \oplus SU(3)  \oplus \RR $  &4\\
\hline
 $A_1 \oplus D(2,1;2)$ & $A_1 \oplus A_1\oplus A_1\oplus A_1$& $SO(2,1)\oplus SU(2)^3 $& 8\\
\hline
  $A(0,3) $ & $U(1) \oplus A_3$ &$ SO(2,4) \oplus \RR $& 8\\
\hline
  $C(3)$& $U(1) \oplus C_2$& $SO(2,3)\oplus SO(2)$ &8\\
\hline
\end{tabular}
\end{center}
\label{table:f4subalg}
\end{table}%

By identifying the $SO(2,p)$ factor in the even part of the sub-superalgebras with the conformal symmetry of a $p$-dimensional defect, one can see that there should be superconformal line defects with $p=1$ preserving 4 and 8 supersymmetries, as well as $p=3$  and $p=4$ dimensional defects preserving 8 supersymmetries. Holographic duals to the $p=1$ defects preserving 8 supersymmetries were explored in \cite{Assel:2012nf,Kaidi:2017bmd}. The goal of the present  paper is to find a realization of the $p=4$ dimensional half-BPS defect in six-dimensional gauged supergravity.

There are two ways to construct defects in $AdS$ spaces. The first is the so-called probe approximation, where one considers a probe brane or string which is embedded in the supergravity background preserving the correct symmetries. In the probe approximation, one considers a small number of branes and neglects the backreaction of the branes on the geometry. One example is found in the holographic dual of four-dimensional ${\cal N}=4$  $SU(N)$ Super Yang-Mills theory, in which a half-BPS Wilson line in the  $l$-th antisymmetric representation is realized as a probe D5-brane  \cite{Gomis:2006sb}. The probe D5-brane has an $AdS_2\times S^4$ worldvolume and $l$ units of electric flux through the $AdS_2$. The worldvolume is embedded inside  the $AdS_5\times S^5$ vacuum of type IIB supergravity and the symmetries of the superconformal defect are realized by the isometries of the embedding. More complicated representations can be achieved by adding additional probe branes. 
  
  Second, one can construct solutions in supergravity without adding branes using a Janus ansatz  \cite{Bak:2003jk}. To obtain a supergravity solution describing a $p$-dimensional defect, one considers a warped product of an $AdS_{p+1}$ factor (potentially combined with other compact manifolds whose isometries realize additional symmetries) over a one- or two-dimensional base manifold. For the example of a half-BPS Wilson line discussed above, the solution \cite{Lunin:2006xr,DHoker:2007mci} is built on a product $AdS_2\times S^2\times S^4$ warped over a two-dimensional Riemann surface. Using the Janus ansatz, solutions corresponding to various half-BPS defects have been constructed in M-theory \cite{DHoker:2008lup,DHoker:2008rje} as well as in type IIB supergravity \cite{DHoker:2007zhm,DHoker:2007hhe}. These solutions are all quite complicated due to the warped product form, the fact that fluxes of the antisymmetric tensor fields are turned on, and the fact that all quantities depend on the two-dimensional base manifold.

A simpler class of supersymmetric Janus solutions corresponding to defects of co-dimension one can be obtained in $(d+1)$-dimensional gauged supergravity theories by considering a metric ansatz of an $AdS_d$ factor warped over a one-dimensional interval. The only other fields which are taken to have a nontrivial dependence on the interval are the scalars.  Supersymmetric Janus solutions  in five- and four-dimensional gauged supergravity were constructed in \cite{Clark:2005te,Suh:2011xc,Bobev:2013yra}. 

The complicated structure of the full type IIB duals of five-dimensional SCFTs makes it very hard to construct the defect solutions which, by the analysis of the sub-superalgebras, should exist. We therefore follow the simplified setting outlined above to construct supersymmetric Janus solutions corresponding to four-dimensional defects in six-dimensional gauged $F(4)$ supergravity.  There is however a price to pay, since it is very difficult to lift lower dimensional gauged supergravity solutions to ten or eleven dimensions,\footnote{Such a lifting has been successfully performed in the case of $\cN = 8$ five-dimension supergravity; see \cite{Suh:2011xc}.} and a clear understanding of the dual CFT is generally not available. However, the simplicity of the system makes finding solutions in the lower dimensional gauged supergravity a worthwhile exercise. 

The plan of the present paper is as follows: In Section \ref{sec2} we review the essential features of matter coupled $F(4)$ gauged supergravity which will be needed in the rest of the paper. In Section \ref{sec3} we describe the Janus ansatz describing a co-dimension one defect and discuss the holographic dictionary for the supergravity scalars. In Section \ref{sec4} we derive the BPS equations by imposing the vanishing of the fermionic supersymmetry transformations for eight of the sixteen supersymmetries. We obtain three coupled nonlinear ordinary differential equations, which are summarized in subsection \ref{sec4.5}. In Section \ref{sec5} we solve the BPS equation numerically and show that the requirement of obtaining smooth and regular solutions reduces the three initial conditions  for a generic solution to a one parameter family. In addition, we use the holographic dictionary to give a field theory interpretation of our solutions. In Section \ref{sec6} we discuss various open questions and avenues for further research. Finally, in the appendices we include some additional information, including an outline of our gamma matrix conventions, a brief review of the pseudo-Majorana condition, and details on the choice of coset representative used in our calculations. 


\section{Matter coupled  $F(4)$ gauged supergravity }
\setcounter{equation}{0}
\label{sec2}

The theory of matter coupled $F(4)$ gauged supergravity was first studied in \cite{Andrianopoli:2001rs,DAuria:2000afl}, with some applications and extensions given in \cite{Karndumri:2016ruc,Karndumri:2012vh}. We model the brief review below on the latter.

\subsection{The bosonic Lagrangian}
The field content of the 6-dimensional supergravity multiplet is 
\bea
(e^a_\mu,\, \psi^A_\mu,\, A^\a_\mu, \,B_{\mu \nu}, \,\chi^A, \,\sigma)
\eea
The field $e^a_\mu$ is the 6-dimensional frame field, with spacetime indices denoted by $\mu, \nu$, and local Lorentz indices denoted by $a,b$.
The field $\psi^A_\mu$ is the gravitino with the index $A ,B= 1,2$ denoting the fundamental representation of the gauged $SU(2)_R$ group. The supergravity multiplet contains four vectors $A_\mu^\a$ labelled by the index $\a = 0, \dots 3$. It will often prove useful to split $\a = (0,r)$ with $r = 1, \dots, 3$ an $SU(2)_R$ adjoint index. Finally, the remaining fields consist of a two-form $B_{\mu \nu}$, a spin-${1 \over 2}$ field $\chi^A$, and the dilaton $\sigma$.

The only matter in the $d=6$, $\cN=2$ theory is the vector multiplet, which has the following field content 
\bea
(A_\mu, \,\lambda_A, \,\phi^\a)^I
\eea
where $I = 1, \dots, n$ labels the distinct matter multiplets included in the theory. The presence of the $n$ new vector fields $A_\mu^I$ allows for the existence of a further gauge group $G_+$ of dimension $\mathrm{dim} \,G_+ = n$, in addition to the gauged $SU(2)_R$ R-symmetry. The presence of this new gauge group contributes an additional parameter to the theory, in the form of a coupling constant $\lambda$. Throughout this section, we will denote the structure constants of the additional gauge group $G_+$ by $C_{I J K}$.  However, these will not play a large role in the rest of the paper, since we will soon restrict to the case of only a single vector multiplet $n=1$, in which case there is no additional gauge group  $G_+$.

More important for our purposes will be the $4n$ scalars $\phi^{\a I}$. Generically in (half-)maximal supergravity, the dynamics of vector scalars is dictated by a non-linear sigma model with target space $G/K$; see e.g. \cite{Samtleben:2008pe}. The group $G$ is the global symmetry group of the theory, while $K$ is the maximal compact subgroup of $G$. As such, in the current case the target space is to be identified with the following coset space, 
\bea
{SO(4,n) \over SO(4) \times SO(n)}
\eea
A convenient way of formulating this non-linear sigma model is to have the scalars $\phi^{\a I}$ parameterize an element $L$ of $G$. This coset representative $L$ is an $(n+4)\times(n+4)$ matrix with matrix elements $L^\Lambda_{\,\,\,\,\Sigma}$, for  $\Lambda, \Sigma = 1, \dots n+4$. We may use this to construct a left-invariant 1-form, 
\bea
L^{-1} d L \in \mathfrak{g}
\eea
where $\mathfrak{g} = \mathrm{Lie}(G)$. To build a $K$-invariant kinetic term from the above, we decompose
\bea
L^{-1} d L = Q + P
\eea
where $Q \in \mathfrak{k} = \mathrm{Lie}(K)$ and $P$ lies in the complement of $\mathfrak{k}$ in $\mathfrak{g}$.  Explicitly, one finds the coset space vielbein forms to be given by, 
\bea
\label{cosetvielbeins}
P^I_{\,\,\,\,\a} = \left(L^{-1}\right)^I_{\,\,\,\,\Lambda} \left(d L^\Lambda_{\,\,\,\,\a} + f^\Lambda_{\,\,\,\,\Gamma \Pi} A^\Gamma L^\Pi_{\,\,\,\,\a} \right)
\eea
where the $f_{\Lambda \Sigma}^{\,\,\,\,\,\,\,\,\Gamma}$ are structure constants of the gauge algebra, i.e.
\bea
[T_\Lambda, T_\Sigma] =  f_{\Lambda \Sigma}^{\,\,\,\,\,\,\,\,\Gamma} \, T_\Gamma
\eea
We may then use $P$ to build the Lagrangian for the vector multiplet scalars as, 
\bea
\cL_{coset} = {1\over 4} e P_{I \a \mu} P^{I \a \mu}
\eea
where $e = \sqrt{|\mathrm{det} \,g|}$ and we've defined $P_\mu^{I \a} = P_i^{I \a} \p_\mu \phi^i,$ for $i = 0, \dots, 4n-1$.  Having expressed the coset space non-linear sigma model as such, we may now write down the full bosonic Lagrangian of the theory. We will be interested in the case in which only the metric and the scalars are non-vanishing. In this case we have 
 \bea
 \label{Lagrangian}
 e^{-1} \cL = -{1 \over 4} R + \p_\mu \sigma \p^\mu \sigma - {1\over 4} P_{I \a \mu} P^{I \a \mu} + V
 \eea
 with the scalar potential $V$ given by 
 \bea
 \label{scalarpot}
V &=& e^{2 \sigma} \left[ {1 \over 36} A^2 + {1 \over 4}B^i B_i + {1 \over 4} (C^I_t C_{I t} + 4 D^I_t D_{I t}) \right] - m^2 e^{-6 \sigma} \cN_{00} 
\no\\
&\vphantom{.}& \hspace{0.5in} + m e^{- 2 \sigma} \left[{2 \over 3} A L_{00} - 2 B^i L_{0 i} \right]
\eea 
The scalar potential features the following quantities, 
\bea
A &=& \eps^{r s t} K_{r s t} \hspace{1 in}\,\,B^r\, = \,\eps^{r s t} K_{s t 0}
\no\\
C_I^t &=& \eps^{t r s} K_{r I s} \hspace{1 in} D_{I t}\,\, = \,\, K_{0 I t}
\eea
The so-called ``boosted structure constants" $K$ are given by,
\bea
K_{r s \a} &=& g \,\eps_{\ell m n} L^{\ell}_{\,\,\,\, r} (L^{-1})_s^{\,\,\,\, m} L^n_{\,\,\,\,\a} + \lambda \,C_{I J K} L^I_{\,\,\,\,r} (L^{-1})_s^{\,\,\,\,J}L^K_{\,\,\,\,\,\a}
\no\\
K_{\a I t} &=& g\, \eps_{\ell m n} L^{\ell}_{\,\,\,\, \a} (L^{-1})_I^{\,\,\,\, m} L^n_{\,\,\,\,t} + \lambda \,C_{M J K} L^M_{\,\,\,\,\,\a} (L^{-1})_I^{\,\,\,\,J}L^K_{\,\,\,\,\,t}
\eea
We remind the reader that $r,s,t = 1, 2, 3$ are obtained from splitting the index $\a$ into a 0 index and an $SU(2)_R$ adjoint index.  Also appearing in the Lagrangian is $\cN_{00}$, which is the $00$ component of the matrix 
\bea
\cN_{\Lambda \Sigma} = L_\Lambda^{\,\,\,\,\a}\left( L^{-1}\right)_{\a \Sigma} -  L_\Lambda^{\,\,\,\,I}\left( L^{-1}\right)_{I \Sigma} 
\eea

\subsection{Supersymmetry variations}
We now move on to the supersymmetry variations of the spinor fields. To begin, we first give some comments on our notation.

 In addition to labelling the 4 vector fields of the supergravity multiplet, the index $\a$ will be used to label Pauli matrices and the identity matrix $\mathds{1}_2$ via 
\bea
\label{sbasis}
\sigma^{\a \,A}_{~~\,B}=(\delta^A_{~B},\sigma^{rA}_{~~B})
\eea
We will make use of the matrix $\g^7$, defined as 
\bea
\g^7=i\g^0\g^1\g^2\g^3\g^4\g^5
\eea
and satisfying $(\g^7)^2=-\mathds{1}$. The indices $A,B$ are raised and lowered with the $SU(2)$ invariant tensor $\varepsilon_{AB}$ as follows:
\bea\label{su2Contr}
T^{\dots A\dots}&=&\varepsilon^{AB}\ \ T^{\dots\ \ \dots}_{\ \ B}\no\\
T_{\dots A\dots}&=&T_{\dots\ \ \dots}^{\ \ B}\ \ \varepsilon_{BA}
\eea
The indices $\a,\b$ are raised/lowered with a Kronecker delta $\delta^\a_\b$, while the indices $I,J$ are raised/lowered with minus the Kronecker delta $-\delta^I_J$. This is because $(\a,I)$ is an index of the global isometry group $SO(4,n)$. 

We may now write the supersymmetry transformations of the fermions as,
\begin{subequations}
\begin{align}
\label{gravitinoVar}
\delta \psi_{\mu A}&\,=\,D_\mu \epsilon_A+S_{AB}\g_\mu \epsilon^B
\\ \label{dilatinoVar} 
\delta\chi_A&\,=\, {i \over 2} \g^\mu \p_\mu \sigma\epsilon_A+N_{AB}\epsilon^B
\\ \label{gauginoVar}
\delta\lambda^I_A&\,=\,iP_{ri}^I\sigma^{r}_{AB}\p_\mu\f^i\g^\mu\epsilon^B-iP_{0i}^I\varepsilon_{AB}\partial_\mu\f^i\g^7\g^\mu\epsilon^B+M^I_{AB}\epsilon^B
 \end{align}
\end{subequations}
where we have defined
\begin{eqnarray}
S_{AB}&=&\!\frac{i}{24}[Ae^{\sigma}\! +\!
6me^{-3\sigma}(L^{-1})_{00}]\varepsilon_{AB}\! -\!
\frac{i}{8}[B_te^{\sigma}-2me^{-3\sigma}(L^{-1})_{t0}]\gamma^7\sigma^t_{AB}\no
\\\no\\
N_{AB}&=&\!\frac{1}{24}[Ae^{\sigma}\! -\!
18me^{-3\sigma}(L^{-1})_{00}]\varepsilon_{AB}\! +\!
\frac{1}{8}[B_te^{\sigma}\! +\! 6me^{-3\sigma}(L^{-1})_{t0}]\gamma^7\sigma^t_{AB}\no\\
\no\\
M^I_{AB}&=&\!(-C^I_{~t}+2i\gamma^7D^I_{~t})e^{\sigma}\sigma^t_{AB}-
2me^{-3\sigma}(L^{-1})^I_{\ \ 0}\g^7\varepsilon_{AB}, \label{mgM}
\end{eqnarray}
In the above, the matrix $\sigma^r_{AB}$ defined as $\sigma^r_{AB}\equiv\sigma^{rC}_{~~B}\varepsilon_{CA}$ is symmetric in $A,B$.

\section{Janus ansatz}
\setcounter{equation}{0}
\label{sec3}

In this section we present the explicit supergravity model that we will be considering in this paper, and discuss the Janus ansatz which will be used to derive the BPS equations in the next section. In addition, we discuss the subtleties which appear in the holographic dictionary for this ansatz.

\subsection{Choice of model}
\label{sec3.1}
For simplicity, we restrict ourselves to the case of $F(4)$ gauged supergravity coupled to a single vector multiplet. The generalization to the case of additional vector multiplets is straightforward. The full scalar manifold in this case is given by
\bea
\cM=\frac{SO(4,1)}{SO(4)}\times SO(1,1)\cong \mathbb{H}^4\times \mathbb{R}^+
\eea
where the $\mathbb{R}^+$ is parameterized by the dilaton $e^\sigma$.  To specify the non-linear sigma model on $\mathbb{H}^4$ concretely, we choose the coset representative
\bea
\label{cosetrep}
L=\prod_{\a=0}^3e^{\phi^\a K^\a}
\eea
where the $K^\a$ are the non-compact generators of $SO(4,1)$. We give the explicit form of these generators in Appendix \ref{cosrep}. With this choice of coset representative, the metric on the scalar manifold descending from (\ref{cosetvielbeins}) is
\bea
G_{ij}=\mathrm{diag} (\cosh^2 \phi^1 \cosh^2 \phi^2 \cosh^2 \phi^3, \cosh^2 \phi^2 \cosh^2 \phi^3, \cosh ^2 \phi^3, 1)
\eea
Having chosen a coset representative, we may calculate explicitly the scalar potential (\ref{scalarpot}), from which we find
\begin{align}\label{scalpot}
V(\sigma,\phi^i) =& g^2 e^{2 \sigma }-\frac{1}{8} m e^{-6 \sigma } \bigg[-32 g e^{4 \sigma } \cosh \phi^0 \cosh \phi^1 \cosh \phi^2 \cosh \phi^3+8 m \cosh ^2\phi^0
\no\\
&+m \sinh ^2 \phi^0 \bigg(-6+8 \cosh ^2\phi^1 \cosh ^2 \phi^2 \cosh (2 \phi^3)+\cosh (2 (\phi^1-\phi^2))\no\\
&+\cosh (2 (\phi^1+\phi^2))+2 \cosh (2 \phi^1)+2 \cosh (2 \phi^2)\bigg)\bigg]
\end{align}
Note that $\phi^0$ is an $SU(2)$ singlet, while the others three scalars $\phi^r$ form an $SU(2)$ triplet. 
The equations of motion follow from the Lagrangian (\ref{Lagrangian}), and are given by
\bea
\label{EOMsigma1}
 \Box~\sigma =\frac12\frac{\delta V}{\delta \sigma}
\eea
for the dilaton and
\bea
\Box~\phi^i+ \p^\mu \phi^i ~\p_\mu (\log G_{ii}) = \frac{2}{G_{ii}}\frac{\delta V}{\delta \phi^i}+\frac{1}{2G_{ii}}\frac{\delta G_{jk}}{\delta \phi^i}\p_\mu \phi^j \p^\mu \phi^k 
\eea
for the vector multiplet scalars. Einstein's equation takes the form,
\bea\label{Eeq}
R_{\mu\nu}=4\p_\mu\sigma\p_\nu\sigma+\p_\mu \phi^i\p_\nu \phi^j G_{ij} + g_{\mu\nu} V 
\eea
For $g=3m$ and vanishing value of the scalar fields, a solution to these equations is the supersymmetric $AdS_6$ vacuum of radius  $l^2 = \(4m^2\)^{-1}$ \cite{DAuria:2000afl,Karndumri:2016ruc}.\footnote{For $g=m$, there is another $AdS_6$ vacuum of radius $l^2=5/(4m^2)$ \cite{Romans:1985tw}. Since this solution violates the BPS equation coming from the dilatino variation, it is non-supersymmetric and provides a dynamical realization of the subalgebra in the first line of Table \ref{table:f4subalg}.}

 In the AdS/CFT correspondence, the masses of supergravity fields are related to the scaling dimensions of their dual operators. For scalars in six dimensions, this relation is
\be
\label{usualthing}
m^2l^2= \Delta(\Delta-5)
\ee
The masses of the scalars can be determined by considering small fluctuations around the $AdS_6$ vacuum. From the scalar potential \eqref{scalpot}, one finds the masses to be \cite{Andrianopoli:2001rs,DAuria:2000afl}
\bea
m_\sigma^2 l^2 = -6 \hspace{0.8 in} m_{\phi^0}^2 l^2 = -4 \hspace{0.8 in} m_{\phi^r}^2 l^2 = -6\,\,,\,\,r=1,2,3
\eea
It follows from \eqref{usualthing} that the $SU(2)$ singlet $\phi^0$ is dual to an operator of dimension $\Delta_{\phi^0} = 4$. On the other hand, $\sigma$ and the $SU(2)$ triplet $\phi^r$ have masses which lie in the double quantization window
\bea
-{25\over 4} \leq m^2 l^2 \leq -{21\over 4}
\eea
where an alternate quantization is possible \cite{Klebanov:1999tb}. If alternate quantization is chosen, these fields correspond to operators of scaling dimension two. However, this choice of boundary condition is inconsistent with supersymmetry since the dual operators are part of conserved current supermultiplets, which constrains the dimension of the scalar operators to be $\Delta_\sigma = \Delta_{\phi^r}=3$ \cite{Ferrara:1998gv}. Therefore we impose standard quantization on $\sigma$ and $\phi^r$.

 We recall that in $AdS$ with Poincar\'e coordinates, the holographic dictionary gives a field theory interpretation to the near-boundary expansions of the scalar fields. In the following, it will be convenient to split the Poincar\'e coordinates as
\be\label{Poincareslice}
ds^2= -{l^2\over \xi^2} \Big( d\xi^2  + dx_\perp^2 -dt^2 + \sum_{i=1}^3 dx_{i} ^2 \Big)
\ee
The coordinate $x_\perp$ measures the distance away from the four-dimensional defect located at $x_\perp=0$. The coordinates  $t, x_i$ with $i=1,2,3$ span the worldvolume of the defect. Then according to the holographic dictionary, the linearized solutions for the scalar fields near the boundary of $AdS_6$ at $\xi=0$ behave as
\be\label{linscala}
\phi \sim \xi^{5-\Delta} \phi_1(t, x)+ \xi^{\Delta} \phi_2(t,x)+ \cdots
\ee
where we identify $\phi_1$ and $\phi_2$ with the source and expectation value, respectively, for the operator dual to the field $\phi$.

\subsection{Janus ansatz}
\label{sec3.2}

In this paper we will be interested in constructing supergravity solutions that describe four-dimensional defects in five-dimensional SCFTs. As discussed in the introduction, the bosonic symmetries of the $AdS_6$ vacuum are $SO(5,2)\times SU(2)_R$, and we aim to construct a four-dimensional superconformal defect which preserves an $SO(4,2)\times U(1)_R$ subgroup of this.\footnote{$U(1)$ is the compact group generated by the Lie algebra $\RR$ listed in the fourth row of Table \ref{table:f4subalg}.} Note that $SO(4,2)$ is the group of isometries of $AdS_5$. Thus to construct such a solution, we utilize an ansatz for the metric which slices the geometry in terms of $AdS_5$ spaces
\bea
\label{DWansatz}
ds^2=-\left(du^2+ e^{2f}d\tilde{s}^2\right)
\eea
where the warp factor $f$ only depends on the slicing coordinate $u$, and the five-dimensional metric $d\tilde s^2$ is given by
\be
d\tilde s^2=  {1\over \zeta^2} \Big(d\zeta^2 -dt^2 +\sum_{i=1}^3{dx_i^2}\Big)
\ee
Furthermore, all scalar fields are taken to depend only on the slicing coordinate $u$.

 As for the breaking of $SU(2)_R$ to $U(1)_R$, this can be achieved by turning on one of the $\phi^r$ which is charged under $SU(2)_R$. In particular, in addition to the uncharged scalars $\sigma$ and $\phi^0$, we choose to switch on the charged scalar $\phi^3$. For simplicity, we keep the other two charged scalars set to zero, i.e. $\phi^1=\phi^2=0$.  It is straightforward to verify that this is a consistent truncation and that it is the most general choice of non-vanishing fields that can (in principle) preserve $SO(4,2)\times U(1)_R$. 
 
On the $AdS$ domain-wall ansatz with the consistent truncation $\phi^1=\phi^2=0$, the equations of motion (\ref{EOMsigma1} - \ref{Eeq}) take the form 
\bea
-(\sigma''+5f'\sigma') &=&\frac12 \frac{\delta V}{\delta\sigma}
\no\\\no\\
-(\phi^{0''}+5f'\phi^{0'}) - \phi^{0'}\(\log \(\cosh^2\phi^3\)\)'&=&\frac{2}{\cosh^2 \phi^3} \frac{\d V}{\d \phi^{0'}}
\no\\\no\\
-(\phi^{3''}+5f'\phi^{3'}) &=&2 \frac{\d V}{\d \phi^{3'}}-\frac12 \sinh(2\phi^3)\(\phi^{0'}\)^2
\no\\\no\\
-5\(f''+\(f'\)^2\) & =& 4(\s')^2 +(\phi^{0'})^2\cosh^2\phi^3 + (\phi^{3'})^2-V
\no\\\no\\
 f''+5 f'^2 + 4 e^{-2 f} &=&    V
\eea

We now briefly review the holographic dictionary for scalar fields in the case of the Janus ansatz. Since we are only interested in identifying the form of the sources and expectation values of the dual operators away from the defect, and not in the calculation of correlation functions, we do not employ the full machinery of holographic renormalization, instead giving a simplified treatment. For a more complete discussion, see e.g.  \cite{Papadimitriou:2004rz,Jensen:2013lxa,Gutperle:2016gfe}

 For the Janus metric ansatz given in (\ref{DWansatz}), the  $AdS_6$ vacuum is obtained by choosing $e^{2f} = \cosh^2 u
$,
\be\label{adsslice}
ds^2=-\left(du^2+ {\cosh^2 u\over \zeta^2} \Big(d\zeta^2 -dt^2 +\sum_{i=1}^3{dx_i^2}\Big)
\right)
\ee
with the boundary of $AdS_6$ at $u\to \pm \infty$.
It is straightforward to verify that a scalar of mass $m$ and dimension $\Delta$  in  the AdS-sliced $AdS_6$   behaves as follows near $u\to +\infty$,
\be\label{expadsa}
\phi \sim \tilde \phi_1 \; e^{-(5-\Delta) u} + \tilde\phi_2 \; e^{-\Delta u} +\cdots 
\ee
It is tempting to identify the constants $\tilde \phi_1 $ and $\tilde\phi_2$ with the source and expectation value, respectively, of the operator dual to $\phi$. However, one has to be careful since this identification works only for asymptotically $AdS_6$ in Poincar\'e coordinates, as in (\ref{linscala}).
For the $AdS_6$ geometry, one can map the Poincar\'e metric \eqref{Poincareslice} to the AdS-sliced metric (\ref{adsslice}) via
\be\label{mapadsads}
x_\perp = \zeta  \tanh u \hspace{0.8 in} \quad \xi = {\zeta\over \cosh u} 
\ee
From (\ref{mapadsads}), it follows that at leading order in $e^u$ we have the following relation between the Poincar\'e coordinates $\xi / x_\perp$ and the coordinate $e^{-u}$,
\be\label{mapasym}
e^{-u} = {\xi\over  2x_\perp}
\ee
Plugging (\ref{mapasym}) into (\ref{expadsa})  gives in the limit $\xi\to 0$
\be\label{adssource}
\phi \sim \tilde \phi_1 \;\left( {\xi \over 2 x_\perp} \right)^{5-\Delta} +  \tilde \phi_2 \;\left( {\xi \over 2 x_\perp} \right)^{\Delta} +\cdots
\ee
Comparing this expression to (\ref{linscala}), it follows that both the source and the expectation value of the operator dual to the scalar field $\phi$ depend on the coordinate $x_\perp$, which is the transverse distance from the location of the four-dimensional defect in the CFT. 

Note that while the map (\ref{mapadsads}) is only exact for pure $AdS$ geometries, the derivation of the position dependence holds even for asymptotically $AdS$ spaces as long as one stays away from the defect, since only the leading behavior in $e^u$ is needed. The construction of a Fefferman-Graham coordinate system which is valid also near the defect is a much more complicated question which we do not address here; see however  \cite{Papadimitriou:2004rz,Jensen:2013lxa}.

\section{BPS equations}
\setcounter{equation}{0}
\label{sec4}

In this section, we use the vanishing of the fermionic supersymmetry variations given in (\ref{gravitinoVar}-\ref{gauginoVar}) to obtain a set of four first-order differential equations. These BPS equations dictate the dynamics of the warp factor $f$ and the three scalars $\sigma$, $\phi^0$, and $\phi^3$ of the theory, while the remaining scalars $\phi^1$, $\phi^2$ are set to zero as per the aforementioned consistent truncation. In addition to these differential equations, we obtain an algebraic constraint that must be satisfied by $f,\sigma,\phi^0$ and $\phi^3$ if they are to give rise to a supersymmetric domain wall solution. The methods used here are similar to those developed in  \cite{Ceresole:2001wi,LopesCardoso:2001rt,LopesCardoso:2002ec} for the study of curved domain walls in five-dimensional gauged supergravity.

\subsection{Projection condition}
We begin by making the following ansatz for supersymmetry projection condition, which respects the pseudo-Majorana condition on $\eps_A$,
\be
\label{projector0}
i\g_5\eps_A=G_0\eps_A-G_3\(\s^3\)^B_{~A}\g^7\eps_B
\ee
This is a consistent projector if
\bea
\label{consistreq}
G_0^2+G_3^2=1
\eea
For the individual $SU(2)$ components, we have
\begin{align}
i\g_5\eps_1=\(G_0-G_3\g^7\)\eps_1 \hspace{1 in} i\g_5\eps_2=\(G_0+G_3\g^7\)\eps_2
\end{align}
It can be checked (using the properties of the gamma matrices given in Appendix \ref{gmc}) that these two conditions on $\eps_1$ and $\eps_2$ are consistent with the pseudo-Majorana condition
\bea
\eps_2=-\g^0\eps_1^*
\eea
Because of this condition, we can pick the eight complex components of $\eps_1$ as the independent spinors. 

By enforcing the projection condition (\ref{projector0}), we have already broken half of the supersymmetries. Thus in the following we will demand  that no more supersymmetries be broken.

\subsection{Dilatino variation}
On the AdS domain wall ansatz (\ref{DWansatz}), the dilatino variation (\ref{dilatinoVar}) reads
\bea
\frac i2\s'\g_5\eps_A=N_0\eps_A+N_3\(\s^3\)^B_{~A}\g^7\eps_B
\eea
where we have defined
\begin{align}
N_0&=-\frac14 \(g\cosh \phi^3 e^\s-3m e^{-3\s}\cosh \phi^0\)\no\\
N_3&=-\frac34 m e^{-3\s}\sinh \phi^0 \sinh \phi^3
\end{align}
Looking at the $A=1$ case, we have that
\bea
\label{projector1}
\s'\(i\g_5\eps_1\)=2\(N_0 + N_3\g^7\)\eps_1
\eea
Using the projector \eqref{projector0} and rearranging, this becomes
\bea
\[\s' G_0-2N_0\]\eps_1-\[\s' G_3+2N_3\]\g^7\eps_1=0
\eea
Since we have already imposed the projection condition (\ref{projector0}), we don't want to impose any further conditions on $\eps_1$. It thus follows that
\bea
 \s'=\frac{2N_0}{G_0}=-\frac{2N_3}{G_3}
\eea
This relation, along with the condition \eqref{consistreq}, determines $G_0$ and $G_3$ in terms of $N_0$ and $N_3$,
\be
G_0=\eta \frac{N_0}{\sqrt{N_0^2+N_3^2}}\hspace{0.7 in}G_3=-\eta\frac{N_3}{\sqrt{N_0^2+N_3^2}}
\ee
where $\eta=\pm1$ .
\subsection{Gaugino variation}
On the AdS domain wall ansatz (\ref{DWansatz}), the gaugino variation (\ref{gauginoVar}) takes the following form
\be
-i\(\cosh \phi^3\(\phi^0\)'\g^7\d^B_{~A}-\(\phi^3\)'\(\s^3\)^B_{~A}\)\g_5\eps_B=M_0\g^7\eps_A+M_3\(\s^3\)^B_{~A}\eps_B
\ee
where we have defined
\begin{align}
&M_0=2m ~e^{-3\s}\cosh \phi^3\sinh \phi^0\no\\
&M_3=-2g ~e^{\s}\sinh \phi^3
\end{align}
From the $A=1$ component, we get
\bea
\label{projector2}
-\(\cosh \phi^3\(\phi^0\)'\g^7-\(\phi^3\)'\)(i\g_5\eps_1)=\(M_0\g^7+M_3\)\eps_1
\eea
Using the projector \eqref{projector0} and rearranging, we find
\bea
&\vphantom{.}&\(\cosh \phi^3\(\phi^0\)' G_3 -\(\phi^3\)'G_0+M_3\)\eps_1 
\no\\
&\vphantom{.}&\hspace{1.3 in}+ \(\cosh \phi^3\(\phi^0\)' G_0 +\(\phi^3\)'G_3+M_0\)\g^7\eps_1=0
\label{gaugino2}
\eea
As before, we don't want to impose any more conditions on $\eps_1$, since we have already imposed the condition (\ref{projector0}). Therefore we obtain a system of two first-order differential equations that can be diagonalized into the following form
\bea
\label{BPSg}
\cosh \phi^3 \(\phi^0\)'&=-\(G_0M_0+G_3M_3\)\no\\
 \(\phi^3\)'&=-\(G_3M_0-G_0M_3\)
\eea
\subsection{Gravitino variation}
On the AdS domain wall ansatz (\ref{DWansatz}), the $A=1$ component of the gravitino variation (\ref{gravitinoVar}) takes the following form
\bea\label{kill1}
D_\mu\eps_1=i \(S_0+S_3\g^7\) \g_\mu\eps_1 
\eea
where we have defined
\begin{align}
S_0&=\frac14 \(g\cosh \phi^3 e^\s+m e^{-3\s}\cosh \phi^0\)\no\\
S_3&=\frac14 m ~e^{-3\s}\sinh \phi^0 \sinh \phi^3\label{S3exp}
\end{align}
and the covariant derivative is
\be
D_\mu\eps=\p_\m\eps+\frac14 \(\omega^{ab}\)_\mu\g_{ab}\eps
\ee
We now consider the integrability of \eqref{kill1} in two different ways, which together will lead to a BPS equation for the warp factor $f$, as well as an algebraic constraint that must be satisfied to have consistent first order equations.

\subsubsection{Integrability condition: first approach}
It follows straightforwardly from \eqref{kill1} that
\be
[D_m,D_n]\eps_1= 2\g_{mn}\(S_0^2+S_3^2\)\eps_1
\ee
The commutator of the covariant derivative gives
\be
[D_m,D_n]\eps_1=\frac14 R_{mnpq}\g^{pq}\eps_1
\ee
On the AdS domain wall ansatz (\ref{DWansatz}), the above component of the Riemann tensor is
\be
R_{mnpq}=\(g_{mp}g_{nq}-g_{mq}g_{np}\)\(\(f'\)^2+e^{-2f}\)
\ee
which then gives the following first-order equation for the warp factor
\bea
\(f'\)^2+e^{-2f}=4\(S_0^2+S_3^2\)
\label{BPSf1}
\eea

\subsubsection{Integrability condition: second approach}
\label{secondint}

We now rewrite \eqref{kill1} in a different manner. First we observe that
\bea
D_m\eps_1=\tilde{D}_m\eps_1-\frac12 f' e^f \tilde{\g}_m\g_5 \eps_1
\eea
where $\tilde{D}_m$ is the covariant derivative on the $AdS_5$ domain wall. Using this equation and our projection condition (\ref{projector0}) for $\eps_1$ results in the following equation 
\bea
\label{gravit1}
\tilde{D}_m\eps_1=-i e^f \tilde{\g}_m\(a-b\g^7\)\eps_1
\eea
where we have defined
\bea
a=\frac12 f' G_0-S_0 \hspace{1 in} b=\frac12 f' G_3-S_3
\eea
Now we impose the integrability of \eqref{gravit1}.  This gives rise to the following formula
\bea
e^{-2f}=\(f'\)^2-4f'\(G_0S_0+G_3S_3\)+4 \(S_0^2+S_3^2\)
\eea
Finally, we make use of the formula for $\(f'\)^2$ in \eqref{BPSf1}, which was obtained from the first integrability condition. One then obtains
\bea
f'=2\(G_0S_0+ G_3S_3\)\label{BPSf2}
\eea
which is the second form of the BPS equation for $f$. Furthermore, \eqref{BPSf1} and \eqref{BPSf2} together give rise to the following algebraic constraint, which must be satisfied when solving the first-order equations
\bea
4 \(S_0^2+S_3^2\)-e^{-2f}=4\(G_0S_0+G_3S_3\)^2\label{constraintrel}
\eea
From this constraint, as well as from the definition of $S_3$ given in (\ref{S3exp}), we see that if we do not switch on either $\phi^0$ or $\phi^3$, then the constraint relation forces $e^{-2f}$ to vanish, and thus for the wall to become flat. Therefore both $\phi^0$ {\it and} $\phi^3$ are needed to support the supersymmetric $AdS$ domain wall solution. The required presence of $\phi^3$ also matches with our expectation from the superalgebra considerations discussed in Section \ref{sec3.2}, since a non-trivial profile for $\phi^3$ breaks the $SU(2)_R$ R-symmetry group to a $U(1)_R$.

\subsection{Summary of first-order equations}
\label{sec4.5}
We now offer a brief review of the results of this section. The first-order equations for the warp factor $f$ and the scalars $\s,\phi^0,\phi^3$ were found to be
\begin{subequations}
\begin{align}
\label{BPSeq1}
f'&=2\(G_0S_0+ G_3S_3\)\\
\label{BPSeq2}
 \s'&=\frac{2N_0}{G_0}\\
 \label{BPSeq3}
\cosh \phi^3 \(\phi^0\)'&=-\(G_0M_0+G_3M_3\)\\
\label{BPSeq4}
 \(\phi^3\)'&=-\(G_3M_0-G_0M_3\)
 \end{align}
\end{subequations}
For consistency, these were required to satisfy the constraint
\be\label{constrainta}
4 \(S_0^2+S_3^2\)-e^{-2f}=4\(G_0S_0+G_3S_3\)^2
\ee
The various functions featured in these equations are defined as
\bea
\label{variousdefs}
S_0&=&\frac14 \(g\cosh \phi^3 e^\s+m e^{-3\s}\cosh \phi^0\)
\no\\
S_3&=&\frac14 m ~e^{-3\s}\sinh \phi^0 \sinh \phi^3
\no\\
N_0&=&-\frac14 \(g\cosh \phi^3 e^\s-3m e^{-3\s}\cosh \phi^0\)
\no\\
N_3&=&-\frac34 m e^{-3\s}\sinh \phi^0 \sinh \phi^3
\no\\
M_0&=&2m ~e^{-3\s}\cosh \phi^3\sinh \phi^0
\no\\
M_3&=&-2g ~e^{\s}\sinh \phi^3
\eea
as well as 
\bea
G_0=\eta \frac{N_0}{\sqrt{N_0^2+N_3^2}} \hspace{0.7 in} G_3=-\eta\frac{N_3}{\sqrt{N_0^2+N_3^2}}\hspace{0.7 in}\eta=\pm 1
\eea

\section{Numerical solutions of the BPS equations }
\setcounter{equation}{0}
\label{sec5}


The BPS equations (\ref{BPSeq1}-\ref{BPSeq4}) are a system of four nonlinear first-order 
ordinary differential equations. Because of the constraint \eqref{constrainta}, there are only three independent functions. An analytic solution of this system is presumably impossible due to the highly nonlinear nature of the equations.
Hence, we will rely on numerical methods to generate solutions.  We note that the methods used in this section are similar to the ones used in \cite{Bobev:2013cja} in a different setting.


\subsection{Asymptotic AdS expansion}

In order to identify proper initial conditions, as well as to obtain a holographic interpretation of solutions to the BPS equations in terms of sources and expectation values of the operators dual to the scalar fields, we perform an expansion in the regime where the scalar fields all decay and the metric approaches an asymptotically $AdS$ form. We make use of the previous discussion in Section \ref{sec3.2}.

We begin by defining an  asymptotic coordinate $z = e^{- u}$, where one side of the asymptotic $AdS$ space is reached by taking $u\to \infty$. Consequently, an asymptotic expansion is an expansion around $z=0$. For an $AdS$ slicing, there is a second asymptotic region given by $u\to -\infty$, which allows for a separate expansion. Note that these two asymptotic regions are not intersecting since they are separated by the region near the defect, where the asymptotic expansions break down.

The coefficients in these expansions may be solved for order by order using the BPS equations. One finds explicitly that all coefficients are determined in terms of only three independent parameters $f_k$, $\a$, and $\b$, in accord with the fact that there are three independent first-order differential equations. The expansions are
\bea
\label{UVexp}
f(z) &=& -\log z +f_k +  \left({1\over 4}e^{-2f_k} - {1\over 16}\a^2\right)  \,z^2 + O(z^4) \no\\
\sigma(z) &=& {3\over 8} \a^2 \,z^2 +  {1\over 4} e^{f_k} \a \b \,z^3 + O(z^4) 
\no\\
\phi^0(z) &=& \a \, z + \left({5\over 4}\a \,e^{-2f_k} - {23\over 48} \a^3\right) \,z^3 + O(z^4)
\no\\
\phi^3(z) &=& e^{-f_k} \a  z^2 + \b \, z^3 + O(z^4) 
\eea
The explicit results for dependence of higher order coefficients on the above three parameters is listed up to $O(z^{8})$ in Appendix \ref{UVIRexp}.  

We recall that $\phi^0$ is dual to an operator of dimension $\Delta=4$, while $\sigma$ and $\phi^3$ are dual to operators of dimension $\Delta=3$. We further recall that (\ref{expadsa}) allows us to identify the sources and expectation values for the dual operators. With this in mind, we can observe that the parameters $f_k$ and $\a$ control the sources for all three operators, whereas the parameter $\b$ controls the expectation value for the operator dual to $\phi^3$.  In addition, the sources and the expectation values have a nontrivial dependence on the distance $x_\perp$ away from the defect, as given by (\ref {adssource}).

The power series (\ref{UVexp}) allows one to set the initial conditions for the numerical integration of the BPS equation at a very small distance away from $z=0$. The space of generic initial conditions is three-dimensional and parameterized by $f_k,\a$, and $\b$. Numerical integration shows that for a generic choice of initial conditions, the solution becomes singular at finite distance in $u$ and hence the supergravity approximation breaks down. Unlike the regular solutions discussed later where a second asymptotic region can be glued smoothly and the solutions describe a defect, for the singular solution there is no second asymptotic region and it is possible to interpret the singular solutions as holographic realizations of boundary conformal field theories \cite{Gutperle:2012hy}. The fact that the geometry is singular and that a clear microscopic picture is lacking where the singularity could for example be replaced by a brane \cite{Fujita:2011fp} limits the usefulness of these solutions. Thus in the next section, we turn towards a study of the conditions required for obtaining non-singular solutions.

\subsection{Regular Janus solutions}

The first thing to note is that obtaining a regular Janus solution will necessarily involve gluing together solutions with opposite values of $\eta$ on either side of the domain wall. Indeed, we may rewrite the BPS equation (\ref{BPSeq2}) for the dilaton $\sigma$ as 
\bea
\label{defsig2}
\sigma' =2 \eta\, \sqrt{N_0^2 + N_3^2}
\eea
from which it is clear that unless $\eta$ changes sign, $\sigma$ will grow indefinitely in one direction. A similar technique of gluing together to obtain regular solutions was also implemented in \cite{Clark:2005te} for the case of a domain wall in five-dimensional gauged supergravity.

When gluing the solutions on either side together, we must ensure smoothness at the gluing point. For example, in the case of $\sigma$, note that 
\bea
\sigma'(u)  \xrightarrow{\eta \rightarrow - \eta} - \sigma'(u)  \xrightarrow{u \rightarrow - u} - \sigma'(-u)
\eea
Smoothness at the origin amounts to the demand that the quantities on the left and right be identified for $u=0$. In other words, we require that 
\bea
\label{smoothness1}
\sigma'(0) = 0
\eea
In the same way, one may show that smoothness requires that both $f'(0)$ and $(\phi^0)'(0)$ vanish as well. It is reassuring that this method of smoothly gluing two solutions together ensures that the warp factor $f(u)$ has a turning point at the location of the domain wall, a characteristic of regular Janus solutions. We make the particular choice of 
\bea
\label{signchoice}
\eta = - \mathrm{sgn}(u)
\eea
so that this turning point is a minimum. This choice of $\eta$ implies further that $\sigma$ and $\phi^0$ have a maximum at the domain wall. 

From the definition of $\sigma$ in (\ref{defsig2}), it is clear that the smoothness condition (\ref{smoothness1}) demands 
\bea
N_0(0) = 0 \hspace{1.5 in} N_3(0) = 0
\eea
Using the definitions of $N_0$ and $N_3$ in (\ref{variousdefs}), this in turn requires that
\bea
\label{sigma02}
\sigma(0) = {1 \over 4}\log\left[{\cosh \phi^0(0) \over \cosh \phi^3(0)}\right]
\eea
as well as 
\bea
\phi^0(0) = 0 \hspace{0.7 in} \mathrm{or}\hspace{0.7 in} \phi^3(0) = 0
\eea
In fact, the case with vanishing $\phi^0(0)$ does not give rise to acceptable solutions. To see this, first consider $\phi^0(0) = 0$ and $\phi^3(0) \neq 0$. From (\ref{sigma02}), it follows that $\sigma(0)<0$. This is unacceptable for the following reason. Recall that for the choice of $\eta$ in (\ref{signchoice}), $\sigma(u)$ obtains a maximum at $u=0$, and furthermore is monotonically decreasing in either direction away from $u=0$. However, since $\sigma(u)$ must vanish at large values of $u$ in order for the geometry to be asymptotically $AdS$, it follows that $\sigma(u)$ must be non-negative. Thus we cannot have $\sigma(0)<0$. For the case of $\phi^0(0) =\phi^3(0) =0$, one finds that $\sigma(0)=0$. The same arguments as above then demand that $\sigma(u)$ vanish identically. In this case, we are simply unable to support a curved domain wall solution. We thus keep $\phi^0(0)$ non-zero, instead choosing $\phi^3(0)=0$. 

It may be checked that for $\phi^3(0)$ vanishing and $\sigma(0)$ as given in (\ref{sigma02}), the smoothness conditions of $f$ and $\phi^0$ are satisfied as well - that is, we automatically have $f'(0)=0$ and $(\phi^0)'(0)=0$. On the other hand, $(\phi^3)'(0)$ does \textit{not} vanish, and thus $\phi^3$ is not smooth at the origin. A simple way to resolve this problem is to switch the sign not only of $\eta$, but also of $\phi^3$ as we cross the domain wall. In this case we have 
\bea
(\phi^3)'(u)  \xrightarrow{\eta \rightarrow - \eta} - (\phi^3)'(u)  \xrightarrow{u \rightarrow - u} - (\phi^3)'(-u) \xrightarrow{\phi^3 \rightarrow - \phi^3}(\phi^3)'(-u)
\eea
Smoothness again requires that we identify the terms on the ends, but we see that it is no longer necessary for $(\phi^3)'(0)$ to vanish. Furthermore, the switch $\phi^3 \rightarrow - \phi^3$ does not affect the discussion of the previous paragraphs, since all of the BPS equations are completely invariant under this transformation. 

Numerical tests indicate that this completes the list of conditions that must be satisfied by solutions which are smooth at the origin and regular at all points. To summarize, we must patch together solutions of opposite $\phi^3$ and $\eta$ on either side of the domain wall, while also enforcing the smoothness constraints 
\bea
f'(0) = 0  \hspace{0.6 in}\sigma'(0) =0  \hspace{0.6 in} (\phi^0)'(0) =0 
\eea
These three constraints are satisfied by demanding that $\phi^3(0)$ vanish and by choosing $\sigma(0)$ to be related to $\phi^0(0)$ as per (\ref{sigma02}). 
The smooth solutions obtained in this way are labeled by a single independent parameter $\phi^0(0)$.

An example of these solutions is shown in Figure \ref{fig1} for the case of $\phi^0(0) = 0.1$. To obtain these plots, we made the particular choice of plotting $-\phi^3$ for $u<0$ and $\phi^3$ for $u>0$. Choosing the opposite sign conventions for $\phi^3$ gives an equally valid result. As mentioned before, we have also required that $\eta$ be given by (\ref{signchoice}), so that $f$ experiences a minimum at the domain wall.

{
\begin{figure}
\centering
\begin{minipage}{.5 \textwidth} 
\centering
\includegraphics[scale=0.85]{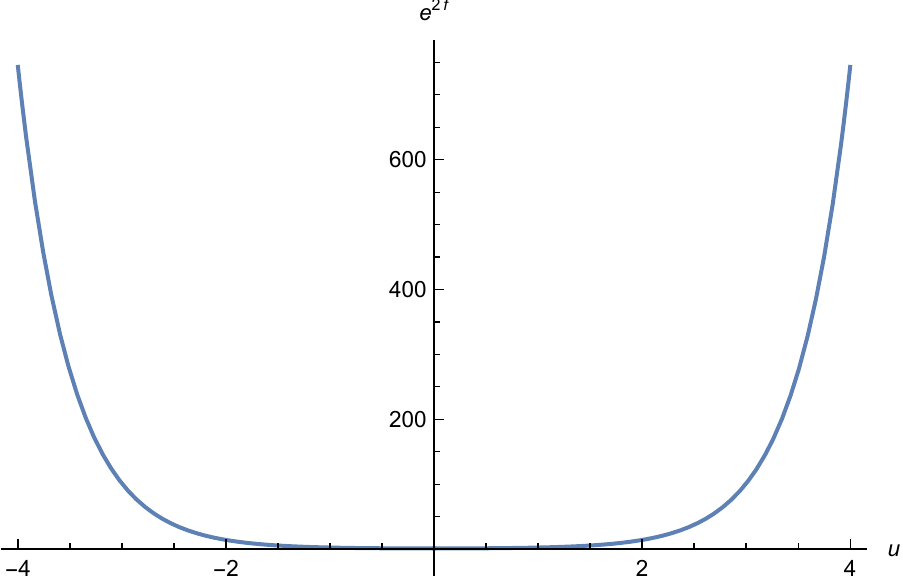}
\end{minipage}%
\begin{minipage}{.5 \textwidth} 
\centering
\includegraphics[scale=0.85]{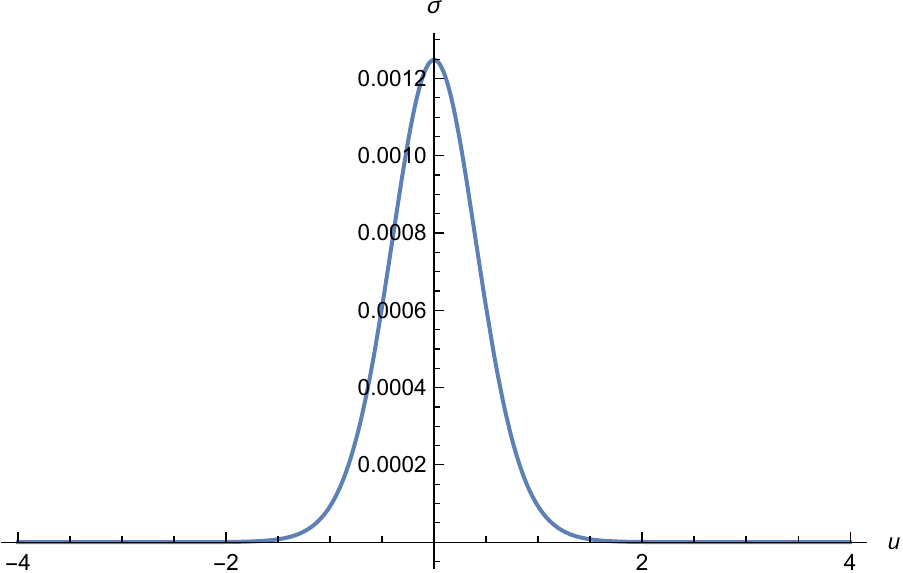}
\end{minipage}%
\newline
\newline
\newline
\centering
\begin{minipage}{.5 \textwidth} 
\centering
\includegraphics[scale=0.85]{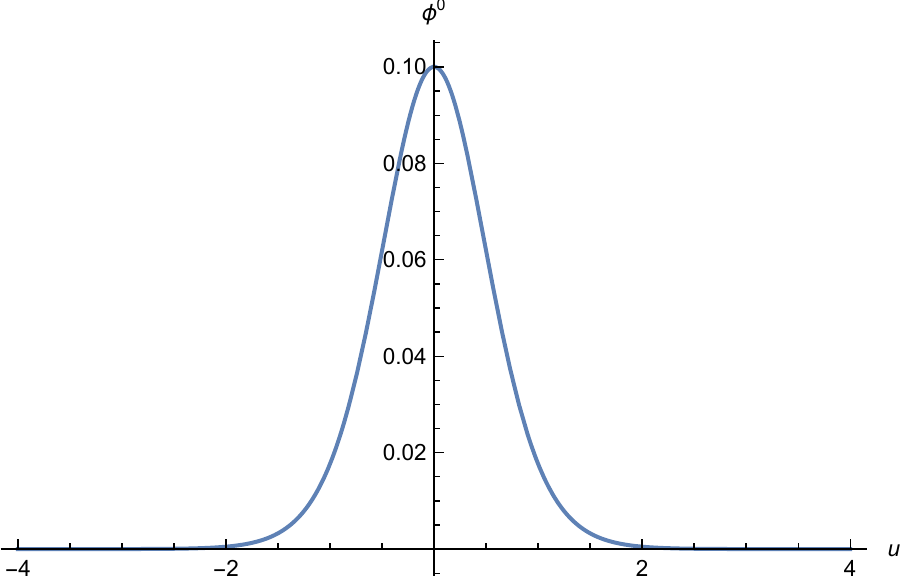}
\end{minipage}%
\begin{minipage}{.5 \textwidth} 
\centering
\includegraphics[scale=0.85]{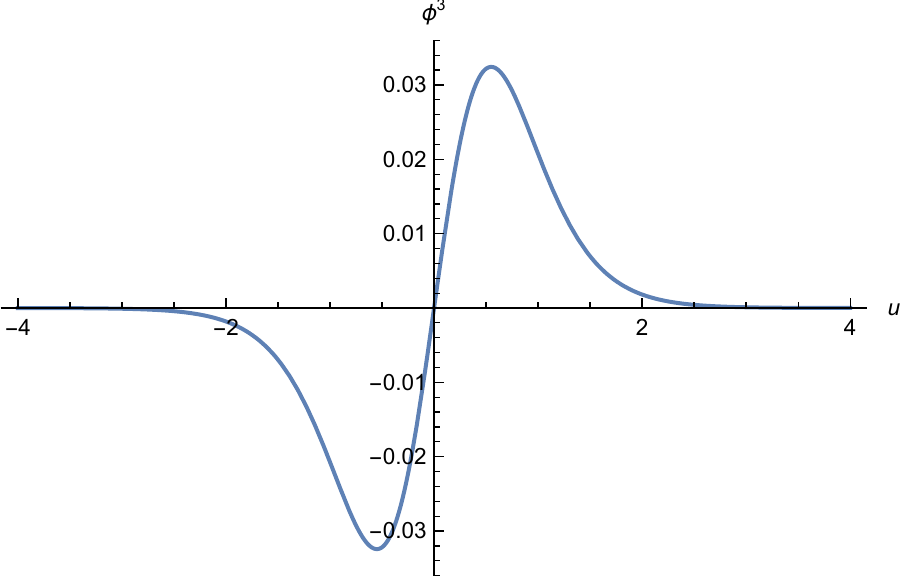}
\end{minipage}%
\caption{Smooth Janus solutions for the four scalar fields. As mentioned in the main text, we take $-\phi^3$ with $\eta = +1$  for $u<0$, and $\phi^3$ with $\eta=-1$ for $u>0$.}
\label{fig1}
\end{figure}
}

\subsection{IR Expansion}
\label{sec5.3}

As seen above, in order to obtain smooth and everywhere regular solutions, we must constrain the values of $\sigma$, $\phi^0$, and $\phi^3$ at $u=0$ via two independent relationships at the origin. In practice though, it is always necessary to impose these constraints not at the origin, but rather at a point very near to the origin, in order to avoid divergences in the numerics.  As such, it will be necessary to understand the behavior of the power series expansions of the scalar fields around $u = 0$. 

As in the UV case, the coefficients in the IR expansions may be solved for order by order via the BPS equations. One finds explicitly that all coefficients are determined in terms of only a single parameter $\phi^0_0$. In particular, one finds 
\bea\label{irexpan}
f(u) &=& -{1\over 4} \log\left[\cosh \phi^0_0 \right] + {1 \over 16}\left(5 + 3 \cosh2 \phi^0_0 \right) \mathrm{sech}^{3 \over2} \phi^0_0\,\, u^2 + O(u^4)
\no\\\no\\
\sigma(u) &=& {1 \over 4} \log\left[\cosh \phi^0_0 \right] - {3 \over 8} {\sinh^2 \phi^0_0 \over \cosh^{3 \over 2} \phi^0_0} \,u^2 + O(u^4)
\no\\\no\\
\phi^0(u) &=& \phi^0_{0}  -2 {\sinh \phi^0_0 \over \cosh^\half \phi^0_0} \, u^2 + O(u^4)
\no\\\no\\
\phi^3(u) &=& {\sinh \phi^0_0 \over \cosh^{3\over4} \phi^0_0} \,\,u -{1\over 48} \left(57 + 31 \cosh2 \phi^0_0\right) {\sinh \phi^0_0 \over \cosh^{9 \over 4} \phi^0_0} \,\, u^3 + O(u^5)
\eea
The expansion coefficients up to $O(u^{8})$ are listed in Appendix \ref{UVIRexp}.

The assumption that a power series expansion around $u=0$ exists implies smoothness at the origin, so we expect to reproduce the smoothness conditions identified in the previous section. Indeed, it is clear from the above that $\phi^3(0) = 0$ and that $\sigma(0)$ is related to $\phi^0(0)$ as per (\ref{sigma02}). Furthermore, we see that $f$, $\sigma$, and $\phi^0$ are even functions of $u$, whereas $\phi^3$ is an odd function of $u$, reversing sign as one crosses the domain wall. These are indeed the requirements for smoothness that were found before.

The existence of this power series ensures that the initial conditions for the numerical integration of the BPS equation can be set not only at $u=0$, but also at a non-zero but sufficiently small distance away from the origin. This justifies the techniques used to obtain the numerical solutions of Figure \ref{fig1}. 
\begin{figure}
\centering
\begin{minipage}{.5 \textwidth} 
\centering
\includegraphics[scale=0.85]{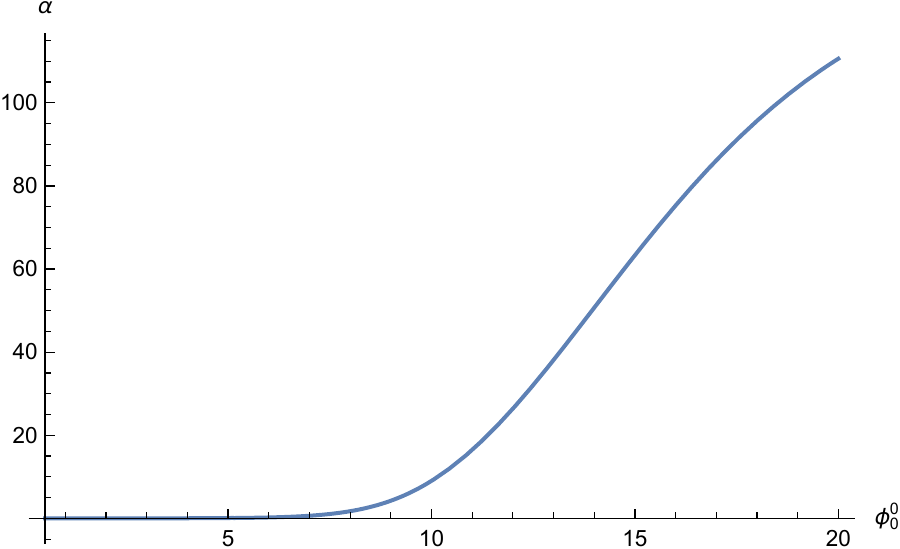}
\end{minipage}%
\begin{minipage}{.5 \textwidth} 
\hspace{0.5 in}
\centering
\includegraphics[scale=0.85]{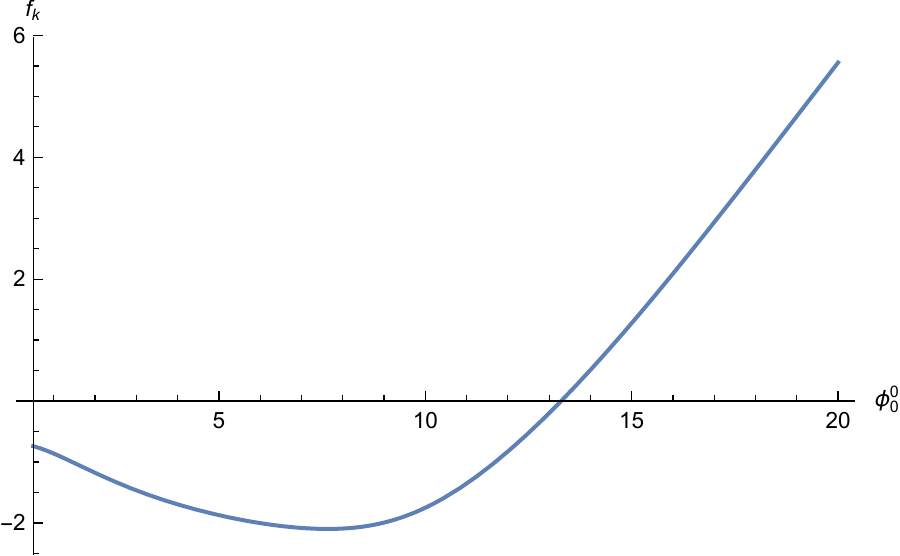}
\end{minipage}%
\newline
\newline
\newline
\begin{minipage}{.5 \textwidth} 
\centering
\includegraphics[scale=0.85]{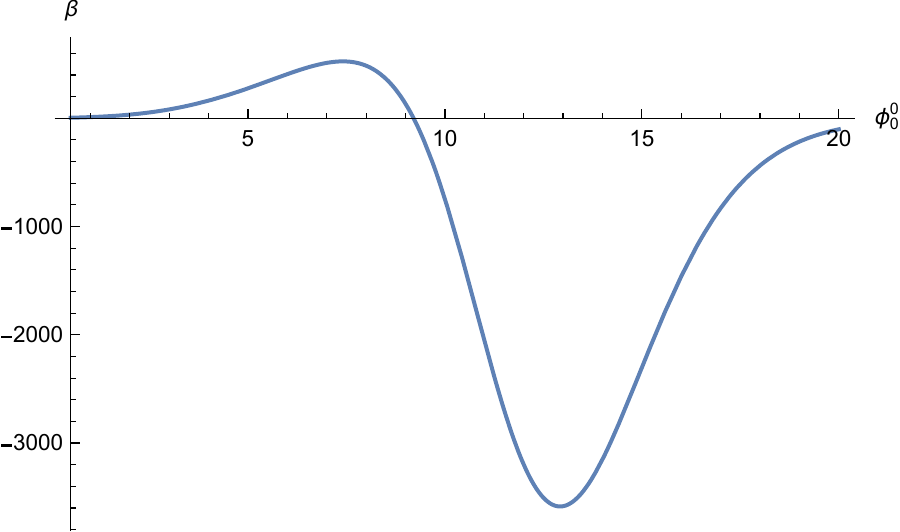}
\end{minipage}%
\caption{Dependences of the UV expansion parameters $f_k$, $\a$, and $\b$ on the IR expansion parameter $\phi^0_0$. }
\label{fig3}
\end{figure}

\begin{figure}
\centering
\begin{minipage}{.5 \textwidth} 
\centering
\includegraphics[scale=0.8]{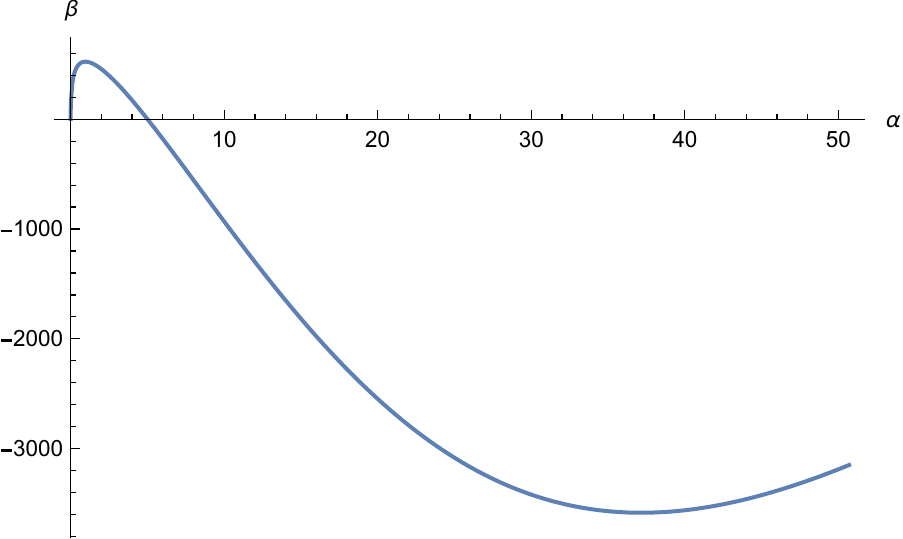}
\end{minipage}%
\begin{minipage}{.5 \textwidth} 
\centering
\includegraphics[scale=0.8]{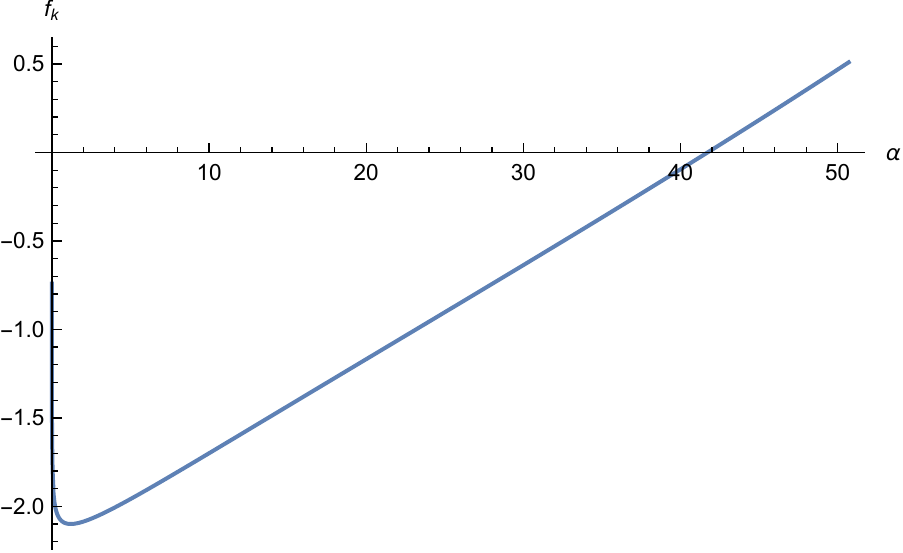}
\end{minipage}%
\caption{Plots of $\b$ vs $\a$ and $f_k$ vs $\a$. The relationships between the three parameters $f_k$, $\a$, and $\b$ may in principle be used to express the UV asymptotic expansion (\ref{UVexp}) in terms of only a single independent parameter.}
\label{fig2}
\end{figure}

\subsection{Relations between the asymptotic parameters for regular solutions}
\label{numericrelations}
The fact that there exists only one free parameter in the IR implies that the three seemingly independent parameters  $f_k$, $\a$, and $\b$ found in the UV must actually be subject to a pair of relationships which constrains them to only a single independent parameter. Though a closed form relationship between the parameters seems difficult to find, numerical relationships are readily obtained. To do so, we first use the IR expansion (\ref{irexpan}) to set the initial conditions for the scalar fields at a location $u_0$ very close to $u=0$. We then use Mathematica to integrate the BPS equations to large values of $u$, and then fit the resulting function to the UV expansion (\ref{UVexp}) to obtain $f_k$, $\alpha,$ and $\beta$ for the particular IR initial condition parameter $\phi_0^0$. Repeating this for various values of $\phi_0^0$ gives the plots of Figure \ref{fig3}.

Note that the CFT interpretation of the IR parameter $\phi_0^0$ is not clear since it is defined at a point deep in the bulk of the spacetime.  It is however straightforward to eliminate $\phi_0^0$ and obtain the functional relationships between pairs of UV parameters.
For example, in Figure \ref{fig2} we plot $\b$ vs. $\a$ as well as $f_k$ vs. $\a$. We can interpret this result as implying that the sources and expectation values for a smooth defect solution are completely determined in terms of the parameter $\alpha$ which controls the source of the operator dual to $\phi^0$.

\section{Discussion }
\setcounter{equation}{0}
\label{sec6}

In this paper we have constructed new solutions of six-dimensional gauged supergravity coupled to one vector multiplet, which provide a holographic realization of a defect in the dual five-dimensional CFT.  The initial conditions of the solution are fine-tuned such that two sides of an $AdS_5$-sliced Janus solution can be glued together smoothly. On the CFT side, the fine-tuning corresponds to a choice of a single parameter which controls the sources and expectation values of the dimension three and four operators dual to the scalars which are turned on. 
The regular solutions correspond to a special class of RG defect where the sources which trigger the RG flow depend on the distance away from the co-dimension one defect. These supergravity solutions are a concrete example of the conclusion derived from the classification of sub-superalgebras of the superconformal algebra $F(4)$ - namely that half-BPS solutions corresponding to co-dimension one superconformal defects must break the $SU(2)$ R-symmetry group to $U(1)$.

There are several possible directions for future research. In the present paper, we constructed only the simplest example of gauged supergravity with a single vector multiplet. Adding more vector multiplets allows for additional symmetries, which on the CFT side are interpreted as global flavor symmetries. It would therefore be interesting to generalize the solution constructed in the present paper to more complicated matter content and gaugings.  

One advantage of constructing solutions in six-dimensional gauged supergravity instead of the ten-dimensional IIB supergravity is that the more complicated warped nature of the ten-dimensional gauged supergravity makes the calculation of holographic observables, such as correlation functions, very challenging. The simpler solutions in six dimensions may be a better starting point. It is an interesting and challenging question whether the solutions found here or generalizations thereof could be be lifted to solutions of ten-dimensional supergravity.

Another interesting point is to understand the nature of the fine-tuning of the initial conditions needed to obtain regular defect solutions instead of  singular solutions. On the CFT side, the vacuum is deformed by position-dependent sources, and it would be very interesting to understand the fine-tuning which leads to the correlation of the strength of the sources and the expectation values of the operators.

As mentioned in Section \ref{sec3.1}, the masses of some of the scalars with nontrivial profiles lie in the window where an alternative quantization is allowed, corresponding to dual operators of dimension two instead of three for $\sigma$ and $\phi^3$. It would be interesting to investigate the interpretation of this alternate quantization for the kind of defect we have constructed. As was mentioned before, such a defect would necessarily be non-supersymmetric.

Finally, analysis of the sub-superalgebras of $F(4)$ in Table \ref{table:f4subalg} implies that in addition to the four-dimensional defect constructed here, there should also exist half-BPS defects with one- and three-dimensional worldvolumes. A study of the former was begun in \cite{Assel:2012nf}, in which Wilson loops for a family of 5d $\cN=1$ SCFTs were examined. The calculation on the gravity side was carried out by embedding probe strings/branes in massive type IIA supergravity backgrounds. Similar calculations were carried out for type IIB in \cite{Kaidi:2017bmd}, in which boundary-anchored probe $(p,q)$-strings were embedded in supergravity backgrounds corresponding to the near-horizon geometry of five-brane webs. In the dual five-dimensional CFT, such probe strings are expected to correspond to Wilson-`t Hooft loops in the fundamental representation \cite{Chen:2006iu}. Seeing as the analysis in both of the above cases was complicated by the warped ten-dimensional geometry, it would be interesting to investigate whether such a holographic defect solution can be more readily constructed in six-dimensional supergravity. Similar statements hold for the three-dimensional defect. In that case, one possibility for a holographic dual would be to consider an $AdS_4\times S^1$ slicing of the six-dimensional spacetime, where the $S^1$ realizes the rotational symmetry around the defect.

We plan to return to these questions in future works.

\section*{ Acknowledgements}

We are grateful to J. Louis and P. Karndumri for a useful correspondence and C. Uhlemann and E. D'Hoker for useful conversations.
The work of M. Gutperle and J. Kaidi is supported in part by the National Science Foundation under grant PHY-16-19926. The work of H. Raj is supported by funding from SISSA. All authors are grateful to the Mani. L.  Bhaumik Institute for Theoretical Physics for support.

\newpage

\newpage

\appendix

\section{Gamma Matrix Conventions}\label{gmc}
\setcounter{equation}{0}

We use the following basis for the gamma matrices.
\begin{align}
\gamma_0 &= \sigma_2 \otimes \mathds{1}_2 \otimes \sigma_3
\no\\
\gamma_1 &= i \sigma_2 \otimes \mathds{1}_2 \otimes \sigma_1
\no\\
\gamma_2 &= i \mathds{1}_2  \otimes \sigma_1  \otimes \sigma_2
\no\\
\gamma_3 &= i \mathds{1}_2 \otimes \sigma_3  \otimes \sigma_2
\no\\
\gamma_4 &= i \sigma_1 \otimes \sigma_2 \otimes \mathds{1}_2
\no\\
\gamma_5 &= i \sigma_3 \otimes \sigma_2 \otimes \mathds{1}_2
\no
\end{align}
With these choices, we have that
\be
\gamma_7 = i\g_0\g_1\g_2\g_3\g_4\g_5=i \sigma_2 \otimes \sigma_2 \otimes \sigma_2
\ee
All gamma matrices are anti-symmetric. All gamma matrices are real except $\g_0$, which is Hermitian. 

 Important for the current work is the fact that in $d=6$, there do not exist spinors $\eps$ satisfying a reality condition of the form 
\bea 
\eps = \gamma_0 \eps^*
\eea
However, if we consider two separate spinors $\eps_1$ and $\eps_2$, a similar condition can be satisfied
\bea
\eps_1 = \gamma_0 \eps_2^* \hspace{1 in} \eps_2 = - \gamma_0 \eps_1^*
\eea
This condition defines a pair of pseudo-Majorana (or symplectic-Majorana) spinors.
\section{Coset Representative}\label{cosrep}
\setcounter{equation}{0}
 
 As stated in \eqref{cosetrep}, we take our coset representative to be given by 
\bea
\label{cosetrep1}
L=\prod_{\a=0}^3e^{\phi^\a K^\a}
\eea
where $K^\a$ are the non-compact generators of $SO(4,1)$. As was done in \cite{Karndumri:2012vh}, we parameterize the group generators by basis elements 
\bea
(e^{x y})_{z w} = \delta_{x z} \delta_{y w} \hspace{0.5 in} w, x, y, z = 1, \dots, 5
\eea
in terms of which the non-compact generators can be written as 
\bea
K^{\a} = e^{\a+1, 5} + e^{5, \a+1}
\eea
\section{UV and IR expansion coefficients}\label{UVIRexp}
\setcounter{equation}{0}

\subsection{UV coefficients}
The UV expansion of (\ref{UVexp}) can be written as
\bea
\label{UVexp2}
f(z) &=& f_k + f_{0} \log z + f_{1} \,z +  f_{2} \,z^2 + f_{3} \,z^3 + f_{4} \,z^4 + \dots
\no\\
\sigma(z) &=& \sigma_{2} \,z^2 + \sigma_{3} \,z^3 + \sigma_{4} \,z^4 + \dots
\no\\
\phi^0(z) &=& \phi^0_{1} \, z + \phi^0_{2} \, z^2 + \phi^0_{3} \,z^3 + \phi^0_{4} \,z^4 + \dots
\no\\
\phi^3(z) &=& \phi^3_{2} \, z^2 + \phi^3_{3}\, z^3 + \phi^3_{4} \,z^4 + \dots
\eea
The coefficients are obtained by solving the BPS equations order by order. All coefficients may be expressed in terms of three free parameters $f_k$, $\a$, and $\b$, as follows
\bea
f_0 &=& - 1
\no\\
f_1 &=& 0
\no\\
f_2 &=& {1 \over 16} \left(4 \,e^{- 2 f_k} - \a^2\right)
\no\\
f_3 &=& 0 
\no\\
f_4 &=& {1 \over 512} \left(-16\, e^{- 4 f_k} - 88 \,e^{-2 f_k} \a^2 + 5 \a^4 \right)
\no\\
f_5 &=& - {1 \over 40} e^{- f_k } \a \b \left(8 + e^{2 f_k} \a^2 \right)
\no\\
f_6 &=& {1 \over 12288}\,e^{- 6 f_k}\, \left( 64 + 4176 \,e^{2 f_k} \a^2-4764 \,e^{4 f_k} \a^4-192\, e^{8 f_k} \a^2 \b^2 - e^{6 f_k}\left(799 \a^6 + 768 \b^2 \right)\right)
\no\\
f_7 &=& - {1 \over 4480}\,e^{-3 f_k} \a \b\, \left(-1568+2116 \,e^{2 f_k} \a^2 + 289 \,e^{4 f_k} \a^4 \right)
\no\\
f_8 &=& {1 \over 262144}\,e^{- 8 f_k}\, \left(-256 - 99072 \,e^{2 f_k} \a^2 + 227808 \,e^{4 f_k} \a^4 + 1152\, e^{10 f_k} \a^4 \b^2 \right.
\no\\
&\vphantom{.}&\hspace{1.4 in} \left. - 16e^{6 f_k} (9773\, \a^6 - 1408\, \b^2) - e^{8 f_k}\a^2 (3513\, \a^6 + 56320\, \b^2)\right)
\no\\
\sigma_2 &=& {3 \over 8} \a^2
\no\\
\sigma_3 &=&  {1 \over 4} e^{f_k} \a \b
\no\\
\sigma_4 &=& {3 \over 64} \a^2 (-4 \,e^{- 2 f_k} + 7\, \a^2)
\no\\
\sigma_5 &=&- {3 \over 64}e^{- f_k} \a \b (-4 + e^{2 f_k} \a^2)
\no\\
\sigma_6 &=&{1 \over 2048} \left(1688\, e^{-2 f_k} \a^4 - 529\, \a^6 + 256\, \b^2 - 16\,e^{- 4 f_k}\a^2(55 + 8 \,e^{6 f_k} \b^2) \right)
\no\\
\sigma_7 &=& - {3 \over 5120} e^{-3 f_k}\a \b \left(1120 - 2016 \,e^{2 f_k} \a^2 + 473\, e^{4 f_k} \a^4 \right)
\no
\eea
\bea
\sigma_8 &=& {3 \over 40960} e^{-6 f_k}\left[9920\, \a^2 - 24880 \,e^{2 f_k} \a^4 + 96\, e^{8 f_k} \a^4 \b^2  \right.
\no\\
&\vphantom{.}&\hspace{1.7 in}\left. - 9 e^{6 f_k} \a^2 (115\, \a^6 - 512 \,\b^2) + 20 e^{4 f_k} (883\, \a^6 - 128 \,\b^2) \right]
\no\\
\phi^0_1 &=& \a
\no\\
\phi^0_2 &=& 0
\no\\
\phi^0_3 &=&{1 \over 48} \left(60 \,e^{-2 f_k} \a - 23\, \a^3 \right)
\no\\
\phi^0_4 &=&{1 \over 4}\b \left(4 \,e^{- f_k} - e^{f_k} \a^2 \right)
\no\\
\phi^0_5 &=&{1 \over 640} \left(-1400 \,e^{-4 f_k}\a + 1540\, e^{-2 f_k} \a^3 - 37\, \a^5 \right)
\no\\
\phi^0_6 &=& {1 \over 40} e^{- 3 f_k} \b \left(-40 + 28\, e^{2 f_k} \a^2 + 11\, e^{4 f_k} \a^4 \right)
\no\\
\phi^0_7 &=&{1 \over 3584} e^{- 6 f_k} \a \left(5880 - 6370\, e^{2 f_k} \a^2 - 875 \,e^{4 f_k} \a^4 + 392 \,e^{8 f_k} \a^2 \b^2 + 2\, e^{6 f_k}(641 \a^6 - 112 \b^2) \right)
\no\\
\phi^0_8 &=& {1 \over 17920} e^{- 5 f_k} \b \left(11200 - 3920 \, e^{2 f_k} \a^2 - 14692 \, e^{4 f_k} \a^4 + 2591\, e^{6 f_k} \a^6 \right)
\no\\\no\\
\phi^3_2 &=& e^{- f_k} \a
\no\\
\phi^3_3 &=& \b
\no\\
\phi^3_4 &=& {1 \over 8} e^{- 3 f_k} \a \left(-20 + 23 \,e^{2 f_k} \a^2 \right)
\no\\
\phi^3_5 &=& {5 \over 16}\,\b \left( -4 \,e^{-2 f_k} + 5 \,\a^2\right)
\no\\
\phi^3_6 &=& {1 \over 768} e^{- 5 f_k} \a \left(1680 - 2408 \,e^{2 f_k} \a^2 + 1167 \,e^{4 f_k} \a^4 + 192 \,e^{6 f_k} \b^2 \right)
\no\\
\phi^3_7 &=& {1 \over 1280} \b \left(1120 \,e^{- 4 f_k} - 528 \, e^{- 2 f_k} \a^2 + 889\, \a^4 \right)
\no\\
\phi^3_8 &=& {1 \over 5120} e^{- 7 f_k} \a \left(-6720 + 1360 \, e^{2 f_k} \a^2 + 6860 \, e^{4 f_k} \a^4 + 384 \,e^{8 f_k} \a^2 \b^2 +3\, e^{6 f_k}\left(455 \a^6 + 1024 \b^2 \right)\right)
\no
\eea

\subsection{IR coefficients}

The IR expansion of (\ref{irexpan}) is given to order $O(u^8)$ by 
\bea
f(u) &=& f_0 + f_1\, u + f_2\, u^2 +f_3 \, u^3 + f_4 \, u^4  + \dots
\no\\
\sigma(u) &=& {1 \over 4} \log\left[\cosh \phi^0(u) \right] + \sigma_2 \,u^2 + \sigma_3\, u^3 + \sigma_4\, u^4 + \dots
\no\\
\phi^0(u) &=& \phi^0_{0}  + \phi^0_1 \,u + \phi^0_2 \, u^2 + \phi^0_3 \, u^3 + \phi^0_4 \, u^4 +\dots
\no\\
\phi^3(u) &=& \phi^3_1 \,u + \phi^3_2 \, u^2 + \phi^3_3 \, u^3 + \phi^3_4 \, u^4 + \dots
\eea
As mentioned in Section \ref{sec5.3}, the scalars $f$, $\sigma$, and $\phi^0$ are even in $u$, and thus all odd coefficients vanish. On the other hand, $\phi^3$ is odd in $u$. The non-zero coefficients are then,
\bea
f_0 &=& -{1\over 4} \log\left[\cosh \phi^0_0 \right]
\no\\
f_2 &=& {1 \over 16}\left(5 + 3 \cosh2 \phi^0_0 \right) \mathrm{sech}^{3 \over2} \phi^0_0
\no\\
f_4 &=&{1 \over 96}\left(4 - 27 \cosh^4 \phi^0_0 + 15 \cosh 2 \phi^0_0 \right)\mathrm{sech}^3 \phi^0_0
\no\\
f_6 &=& {1\over 11796480} \left[816480 \cosh^{3 \over 2} \phi^0_0 + \mathrm{sech}^{9 \over 2} \phi^0_0 \left( 347026 - 1174629 \cosh 2 \phi^0_0 \right. \right.
\no\\
&\vphantom{.}&\hspace{2.5in} \left. \left. +187326 \cosh 4 \phi^0_0 + 85941 \cosh 6 \phi^0_0 \right)\right]
\no\\
f_8 &=& -{1 \over 10321920} \mathrm{sech}^{6}\phi^0_0\left(726755 - 1032504 \cosh 2 \phi^0_0 + 232092 \cosh 4 \phi^0_0\right. 
\no\\
&\vphantom{.}&\hspace{2.5in} \left.  + 113400 \cosh 6 \phi^0_0 + 29889 \cosh 8 \phi^0_0 \right)
\no\\
\sigma_2 &=& {1 \over 8} \mathrm{sech}^{3 \over 2}\phi^0_0 \sinh^2 \phi^0_0
\no\\
\sigma_4 &=&-{11 \over 192}\left(5 + 3 \cosh 2 \phi^0_0 \right) \mathrm{sech} \phi^0_0\, \tanh^2 \phi^0_0
\no\\
\sigma_6 &=&{1 \over 92160} \left(38555 + 36156 \cosh 2 \phi^0_0 + 10281 \cosh 4 \phi^0_0 \right) \mathrm{sech}^{9 \over 2} \phi^0_0 \, \sinh^2 \phi^0_0
\no\\
\sigma_8 &=& - {1 \over 2580480}\mathrm{sech}^{4}\phi^0_0\, \tanh^2 \phi^0_0 \left( 1195346+1317981 \cosh 2 \phi^0_0  \right.
\no\\
&\vphantom{.}&\hspace{2.5in} \left.+ 784446 \cosh 4 \phi^0_0 + 184851 \cosh 6 \phi^0_0 \right)
\no\\
\phi^0_2 &=& - 2\, \mathrm{sech}^{\half}\phi^0_0 \, \sinh \phi^0_0
\no\\
\phi^0_4 &=&{1 \over 6} \left(15 \sinh \phi^0_0 - \mathrm{sech}\phi^0_0 \, \tanh \phi^0_0 \right)
\no\\
\phi^0_6 &=& - {1 \over 5760}\mathrm{sech}^{7 \over 2}\phi^0_0 \, \sinh \phi^0_0 \left(2195 + 6396 \cosh 2 \phi^0_0 + 3441 \cosh 4 \phi^0_0 \right)
\no\\
\phi^0_8 &=& {1 \over 161280} \mathrm{sech}^5 \phi^0_0 \sinh \phi^0_0 \left(- 14170 + 84987 \cosh 2 \phi^0_0 \right. 
\no\\
&\vphantom{.}& \hspace{2.5 in} + 132522 \cosh 4 \phi^0_0 + 53685 \cosh 6 \phi^0_0\left. \right)
\no\\
\phi^3_1 &=& \mathrm{sech}^{3 \over 4} \phi^0_0 \, \sinh \phi^0_0
\no\\
\phi^3_3 &=&- {1 \over 48} \mathrm{sech}^{9 \over 4} \phi^0_0 \sinh \phi^0_0 \left(57 + 31 \cosh 2 \phi^0_0 \right)
\no\\
\phi^3_5 &=&{1 \over 30720} \mathrm{sech}^{15 \over 4} \phi^0_0 \left( 19606 \sinh \phi^0_0 + 6725 \sinh 3 \phi^0_0 + 4383 \sinh 5 \phi^0_0\right)
\no\\
\phi^3_7 &=& - {1 \over 20643840} \mathrm{sech}^{21 \over 4} \phi^0_0 \left(10980773 \sinh \phi^0_0 - 1308275 \sinh 3 \phi^0_0 \right.
\no\\
&\vphantom{.}& \hspace{2.5 in}\left. + 3667761 \sinh 5 \phi^0_0 + 1434249 \sinh 7 \phi^0_0 \right)
\no\\\no
\eea

\end{document}